\documentclass[%
 reprint,
superscriptaddress,
 amsmath,amssymb,
 aps,
]{revtex4-2}

\usepackage{graphicx}
\usepackage{dcolumn}
\usepackage{bm}

\begin{document}

\title{The Intimate Relationship between Cavitation and Fracture}

\author{Shabnam Raayai-Ardakani}
\author{Darla Rachelle Earl}
\affiliation{ Massachusetts Institute of Technology, Department of Civil and Environmental Engineering, Cambridge, MA, 02139, USA}
\author{Tal Cohen}
\email{Corresponding author: talco@mit.edu}
\affiliation{ Massachusetts Institute of Technology, Department of Civil and Environmental Engineering, Cambridge, MA, 02139, USA}
\affiliation{ Massachusetts Institute of Technology, Department of Mechanical Engineering, Cambridge, MA, 02139, USA}

\date{\today}

\begin{abstract}
Nearly three decades ago, the field of mechanics was cautioned of the obscure nature of cavitation processes in soft materials [Gent, A.N., 1990. Cavitation in rubber: a cautionary tale. Rubber Chemistry and Technology, 63(3)]. Since then, the debate on the mechanisms that drive this failure process is ongoing. Using a high precision volume controlled cavity expansion procedure, this paper reveals the intimate relationship between cavitation and fracture. Combining a Griffith inspired formulation for crack propagation, and a Gent inspired formulation for cavity expansion,  we show that despite the apparent complexity of the fracture patterns, the pressure-volume response follows a predictable path. In contrast to available studies, both the model and our experiments are able to track the entire process including the unstable branch, by controlling the volume of the cavity. Moreover, this minimal theoretical framework is able to explain the ambiguity in previous experiments by revealing the presence of metastable states  that can lead to first order transitions at onset of fracture. The agreement between the simple theory and all of the experimental results conducted in PDMS samples with shear moduli in the range of 25-246 [kPa], confirms that cavitation and fracture work together in  driving the expansion process. Through this study we also  determine the fracture energy of PDMS and show its significant dependence on  strain stiffening. 
\end{abstract}

                              
\maketitle

\noindent When pushed to its extremes, a solid will usually fail by one of two mechanisms; cavitation or fracture. The former refers to the spontaneous growth of pre-existing defects within the body, while the latter is characterized by splitting the material to form new surface area. Fracture is something that we experience often, from the shattering of a window or a cellphone screen and to the splitting of a piece of fruit. Cavitation, on the other hand, is not as mundane, nonetheless it is  ubiquitous. Theories of cavitation have been used to explain extreme phenomena that occur over various length scales; from the formation of craters by meteorite impact \cite{goodier1964mechanics,melosh2005planetary}, volcanic eruptions \cite{volcano1, volcano2}, underground explosions \cite{chadwick1964mechanics,chyba19931908,cohen2010shock,cohen2013hypervelocity}, and penetration phenomena \cite{forrestal1997spherical,sobeski2015review};  to  the exposure of seeds in ripening crop (see Supplementary Information S2), and to  morphogenesis processes  that occur in crucial stages of embryonic development \cite{hoijman2015mitotic,navis2016pulling}. Moreover, cavitation  is employed in medical applications to enhance drug delivery, and to treat cancer \cite{coussios2008applications, wood2015review, stride2010cavitation}, it has been indicated as a primary mechanism of damage in traumatic brain injury \cite{el2008biomechanics,salzar2017experimental,goeller2012investigation}, and has been shown to provide a means for probing of local material properties  in soft and biological materials \cite{zimberlin2007cavitation,crosby2011blowing,cohen2015dynamic,raayai2019volume,estrada2018high}.

Although cavitation and fracture have been traditionally considered separately, it is well known that one can lead to the other. 
In ductile metals, growth and coalescence of voids is commonly accepted as the primary mechanism of fracture \cite{rogers1960tensile, mcclintock1968criterion,tvergaard1982ductile, ashby1989flow, tvergaard1984analysis}. In rubbers and highly deformable materials, however, there is no agreement as to which mechanism is activated at every instance \cite{gent1990cavitation} and the debate is ongoing \cite{poulain2017damage}. On the one hand, an extensive body of literature has been devoted to investigation of cavitation as a purely elastic phenomenon  \cite{horgan1995cavitation, ball1982discontinuous,stringfellow1989cavitation, hang1992cavitation,gent1959internal}. On the other hand, the same phenomenon has been consistently considered in the literature as a fracture process \cite{lindsey1967triaxial,lin2004cavity,lefevre2015cavitation, kundu2009cavitation, hutchens2016elastic,gent1991fracture}. In a recent study, Poulain et al. \cite{poulain2017damage}, were able to observe the process at high spatial and temporal resolutions and conclude that ``internal damage in soft material should be viewed as a fracture phenomenon''.

While the general debate in the literature revolves around the question: \textit{Is it cavitation or fracture?}, in this work we show that the two mechanisms are intimately coupled and inseparable. Additionally we show that the fully coupled behaviour can be captured analytically without employing any additional fitting parameters. 

A major challenge in experimentally studying the formation of internal cavities/cracks is in controlling the process while clearly capturing the form of the cavity/crack in the bulk. Recent advances have been pushing the experimental capability into high speed fracture phenomena \cite{MILNER2019} and employing phase separation for identification of the crack initiation \cite{kim2018scale}. Additionally, these processes are known to be highly sensitive to initial imperfections \cite{gent1969surface,gent1991fracture,lopez2011cavitation} or can only identify the onset of the fracture \cite{kim2018scale}. In available studies \cite{poulain2017damage,hutchens2016elastic}, neither the volume of the cavity nor the applied load have been controlled. Hence, formation of the cavity/crack occurs abruptly.

In the present work, by controlling the volume of the cavity/crack as it expands within a transparent PDMS sample, we are able to obtain precise measurements of pressure $(p)$ versus volume $(V)$ throughout the process not only at the initiation of the crack but also throughout the crack propagation. This is achieved by our custom designed experimental setup \cite{raayai2019volume}, which  functions as a volume controlled syringe pump (see schematic illustration in Figure \ref{fig:patterns}). 
We fabricate cuboid PDMS samples of various shear moduli, in the range of $25-246$  $\rm[kPa]$, as achieved by varying the base to cross-linking agent ratios, and we inflate a cavity/crack at the tip of the needle at a constant volumetric rate \footnote{Note that the diameter of the needle is chosen to be sufficiently large so that surface tension effects can be neglected. The rate of inflation is sufficiently slow such that the process is quasi-static and no rate dependence is observed.}. The fabrication protocol and experimental procedure are detailed in Supplementary Information (S1). 

Even if viewed solely as a fracture process, the expansion behaviour observed in our experiments, is  extremely complex. To demonstrate this, after performing the volume controlled experiment to a predefined volume, we drain the injected fluid and replace it with a fast curing polymeric material (polyvinyl siloxane) and `freeze' the shape of our cavities/cracks as described in the Supplementary Information (S1). 

Images of representative cavities/cracks that have been `frozen' using this method are shown in Fig. \ref{fig:patterns} for samples of different shear moduli $(\mu)$. It is apparent that the intricate patterns that form vary considerably between the specimens and even within the same cavity/crack, they can have several protrusions, and regions of different surface roughness. These differences may be attributed to the presence of initial imperfections created in the injection process. Nonetheless, if considering only the aspect ratio of the cavities/cracks, it is observed that stiffer specimens take forms of higher aspect ratio (additional images can be found in the Supplementary Information S3).

\begin{figure}[ht]
\centering
\includegraphics[width=\linewidth]{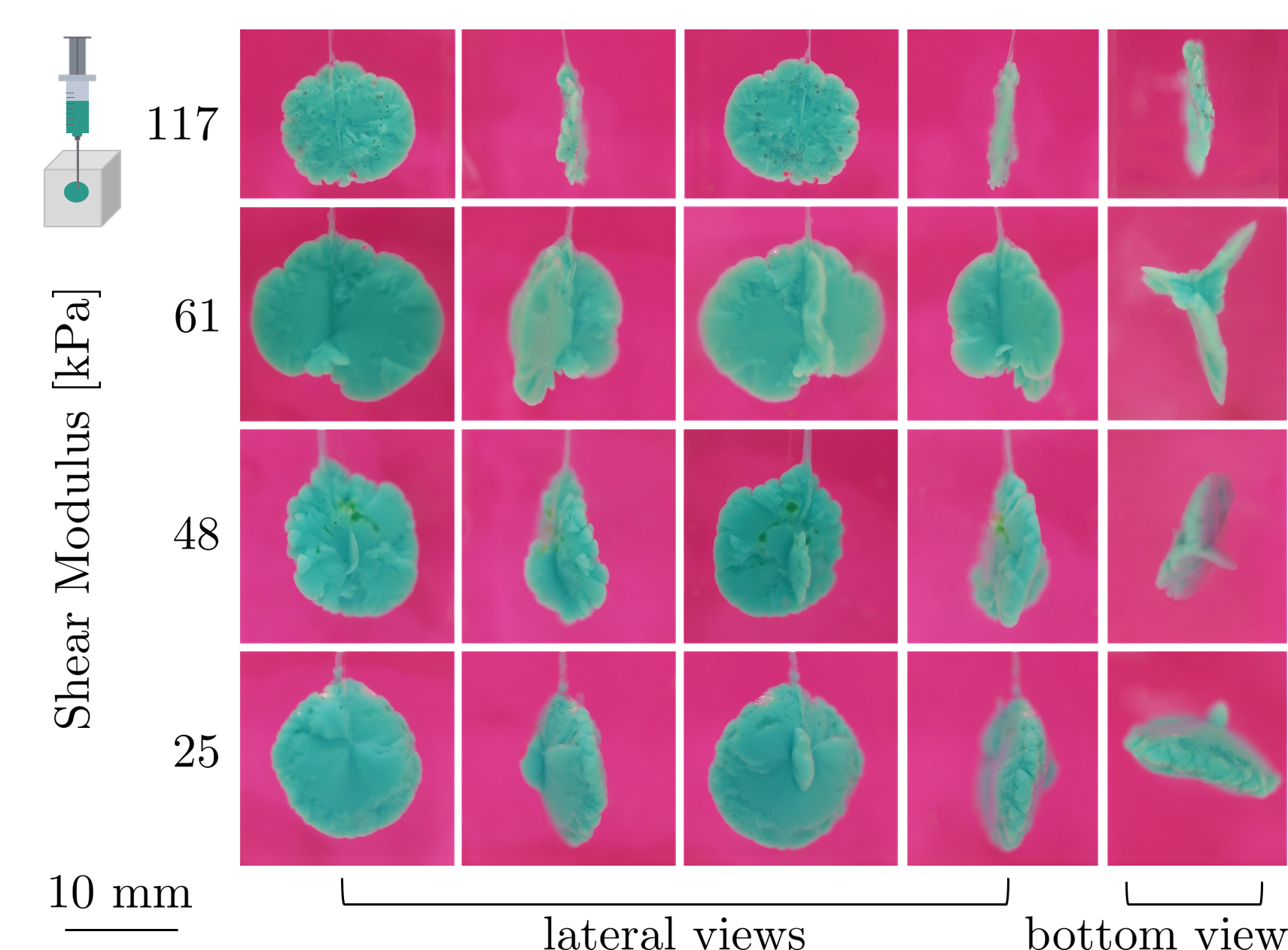}
\caption{Lateral and bottom views of the expanded cavities/cracks in transparent PDMS samples with various shear moduli. The cavities were filled with polyvinyl siloxane (PVS) to visualize the texture of the fracture surfaces. The background colour is for ease of visualization. A schematic of the experimental method is shown on the top left hand side.}
\label{fig:patterns}
\end{figure}

Despite the intricacy observed in Fig.\ \ref{fig:patterns}, the  pressure-volume response exhibits astonishing order. In Fig.\ \ref{curves} we show representative results for three different values of the shear moduli (additional curves can be found in the Supplementary Information S4). The pressure is shown as a function of the effective stretch, $\lambda=(V/V_0)^{1/3}=a_0/A_0$, where $V_0$ is the volume of the initial defect, $A_0$ is the effective radius of the initial defect, and $a_0$ is the effective radius of the cavity/crack (see Fig. \ref{UF}a for illustration, and the Supplementary Information S1 for experimental details). This effective stretch provides a meaningful dimensionless measure for the cavity expansion that is independent of the initial volume. The response can be divided into three stages. At the initial stage of inflation, all of the experimental curves follow a similar path,  then departure from that path is apparent at a peak value, but does not occur at the same point for each test. Nonetheless, as expansion progresses, the curves seem to resume a shared path. In this process the peak value of cavity pressure represents the stability limit. If the volume is not constrained, any pressure above this peak value would result in complete rupture of the specimen. As observed in Fig. \ref{curves}, for stiffer samples the peak value increases and is followed by a steeper decline.  

\begin{figure*}[!ht]
\centering
\includegraphics[width=\textwidth]{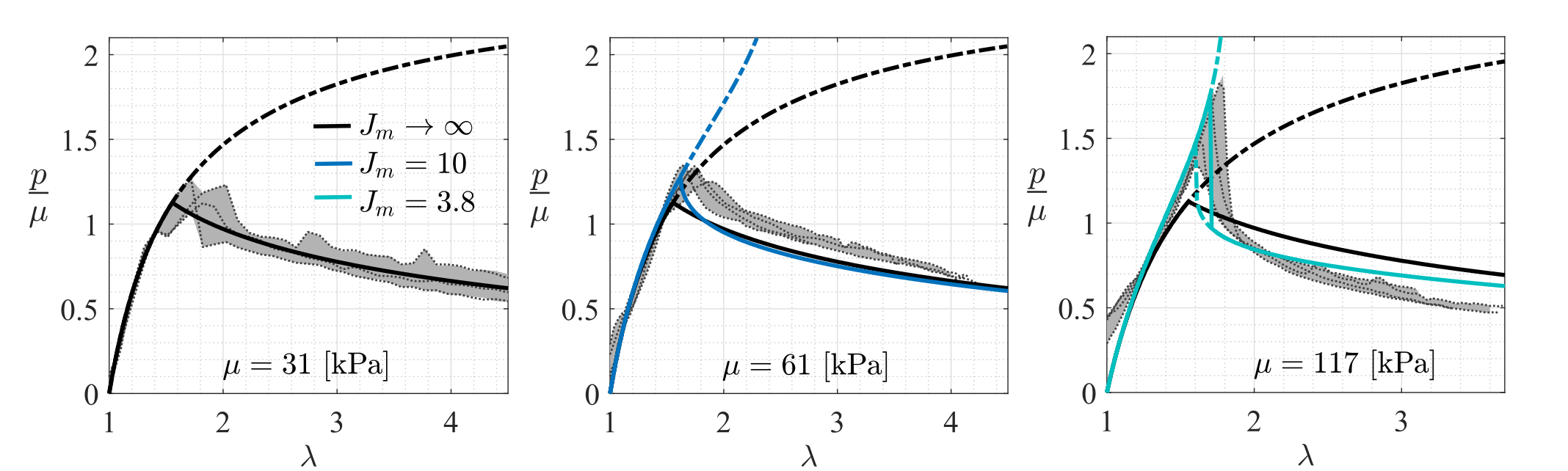}
\caption{Normalized pressure, measured inside cavities of PDMS samples with different shear moduli shown as a function of the effective stretch. Experimental curves are shown by the thin dotted lines. At least three tests were conducted for each material. The solid and dashed-dotted lines represent the predicted response using the present model, with the dashed-dotted lines representing metastable paths.}
\label{curves}
\end{figure*}

Although the experimental observations indicate that this process is highly sensitive to initial imperfections, the conforming of the experimental curves to similar trends, in all but an intermediate range, suggests that a theoretical model can potentially aid in determining which mechanism drives this process (cavitation, fracture, or both).

Complementary to the previous analysis of fracture initiation \cite{Williams1965,gent1991fracture,lin2004cavity}, here we assume that the deformation of the sample can be divided into two distinct zones that border at an effective radius that is denoted by $A$ in the undeformed configuration and  by $a$ in the deformed configuration (Fig. \ref{UF}a). The deformation in the external region is assumed to be purely elastic and spherically symmetric. Limiting our discussion to incompressible materials, the kinematics of the deformation in this region is fully defined as a function of the circumferential stretch at the boundary $\lambda_a=a/A$. Considering bodies of finite dimension  we also write the circumferential stretch at the external boundary as $\lambda_b={b/B}$ where the effective external  radius of the body is $B$ in the undeformed and $b$ in the deformed configuration (it will be shown later that the dimension of the body does not effect the response). By integrating the elastic energy density over the volume of the body , the total elastic energy can  be written, after some algebra, as in \cite{raayai2019volume}
\begin{equation}\label{UE}
    U_E=4\pi \mu A^3(1-\lambda_a^3)\int\displaylimits_{\lambda_a}^{\lambda_b}\frac{\bar W( \lambda_\theta)}{( 1-\lambda_\theta^3)^2}\lambda_\theta^2{\rm d} \lambda_\theta
\end{equation}
where, for the considered deformation, the elastic energy density (per unit volume) of an arbitrary hyperelastic material  $W=\mu \bar W(\lambda_\theta)$ depends only on the shear modulus ($\mu$), and the local value of the circumferential stretch ($\lambda_\theta=r/R$), where $r$ and $R$ represent the radial coordinate in the current and reference configurations, respectively (Fig. \ref{UF}a).

The internal region $R\in[A_0,A]$ is assumed to undergo a fracture process. To estimate the energy invested in creating surface area within the fractured region, we employ an augmented Griffith \cite{griffith1921vi} type approach to account for the spherical nature of the deformation and the effect of strain stiffening
\begin{equation}
    U_F=2\pi(A^2-A_0^2)G_cf(\lambda_a).
\end{equation}
Here the the geometric multiplier $2\pi (A^2-A_0^2)$ provides a measure for the affected surface area, and $G_c$ is an \textit{effective measure} of the fracture energy per unit area. Note that $G_c$ cannot be directly compared to the reported values in the literature; since  the crack is free to take any form, it does not correspond to a predefined unit area. Here, we extend this energy based analysis not only to determine the point of fracture initiation but also to predict the pressure response during the propagation of the crack. This extends the approach used in \cite{Williams1965,gent1991fracture,lin2004cavity}, by controlling the cavity size to access the unstable branch. To account for the possible influence of strain stiffening on the energetic cost of advancing the crack (as it has also been previously discussed \cite{gent1991fracture}), we include an additional (dimensionless) function $f(\lambda_a)$, which will be discussed later in further detail.

The total energy invested in the expansion process is the sum of the two energies $U=U_E+U_F$ and takes the dimensionless form
\begin{equation}\label{totalU}
    \frac{U}{2\pi \mu A_0^3}=\left(\frac{A}{A_0}\right)^3g(\lambda_a,\lambda_b)+\varphi\left(\left(\frac{A}{A_0}\right)^2-1\right)f(\lambda_a)
\end{equation}
where $\varphi=G_c/(\mu A_0)$ is a dimensionless measure of fracture energy, and 
\begin{equation}
    g(\lambda_a,\lambda_b)=2(1-\lambda_a^3)\int\displaylimits_{\lambda_a}^{\lambda_b}\frac{\bar W( \lambda_\theta)}{( 1-\lambda_\theta^3)^2}\lambda_\theta^2{\rm d} \lambda_\theta
\end{equation}

Note that due to incompressibility, for a given imposed expansion $a_0$, the stretches $\lambda_a$ and $\lambda_b$ depend on the initial dimensions of the body $A_0$, $B$, and on the location of the boundary  $A \in [A_0 \ \ B].$  
It is expected that for a given imposed expansion $a_0$, the equilibrium location of the boundary $(A)$ will minimize the total energy, such that ${\partial U}/{\partial A}=0$ and ${\partial^2 U}/{\partial A^2}>0$. Additional extrema may appear at the physical boundaries. An extramum at $A = A_0$ corresponds to the purely elastic solution, while an extramum at $A = B$ corresponds to complete failure.
  
\begin{figure*}[ht]
\centering
\includegraphics[width=\textwidth]{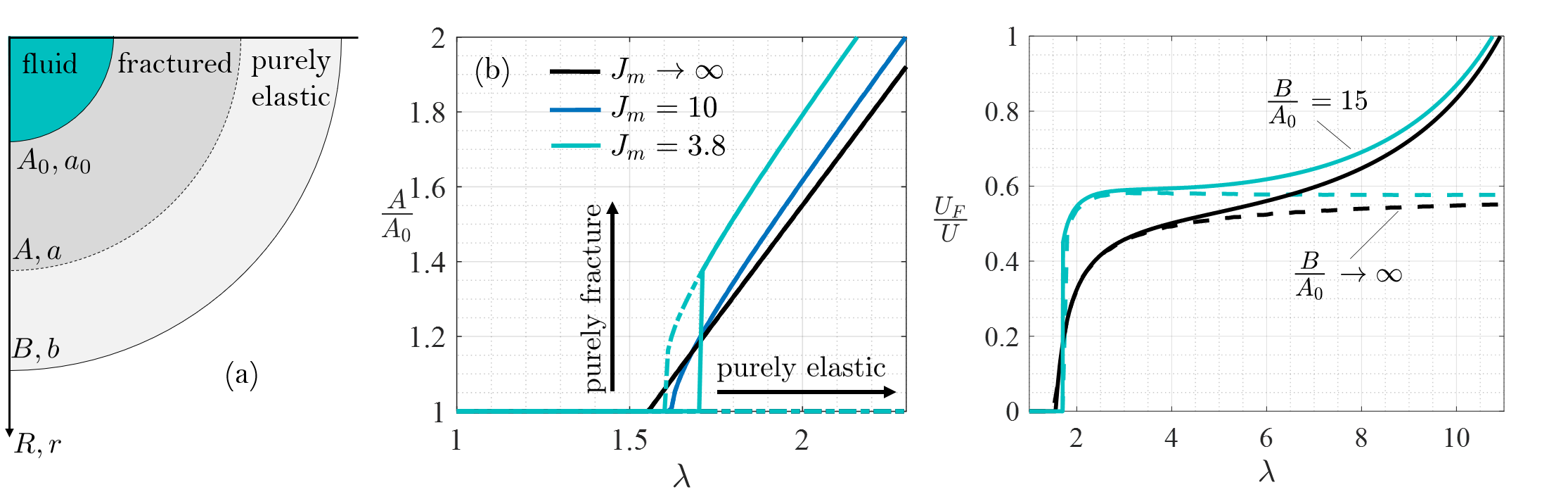}
\caption{(a) Illustration of theoretical spherically symmetric expansion field. The initial and current effective radius of the inflated cavity/crack is denoted by $A_0$ and $a_0$, respectively. Similarly, the external radius is denoted by $B$ and $b$ and the boundary between the fractured region and the purely elastic region is denoted by $A$ and $a$, respectively. (b) Phase diagram of the cavity expansion process shown for the different levels of material stiffening and calculated for $B/A_0=15$. The solid and dashed-dotted lines represent the predicted response using the present model, with the dashed-dotted lines representing metastable paths. (c) Fraction of fracture energy throughout the expansion process  shown for different ratios $B/A_0$ and for different levels of material stiffening.}
\label{UF}
\end{figure*}

Assuming that the fractured material is no longer capable of resisting the deformation, the cavity pressure can be estimated by considering only the elastic region and can be obtained by the derivative $p=-\partial U_E/\partial V$, which can be simplified to the form
\begin{equation}\label{pE}
    \frac{p}{\mu}=\int\displaylimits_{\lambda_a}^{\lambda_b}\frac{\bar W'( \lambda_\theta)}{ 1-\lambda_\theta^3}{\rm d} \lambda_\theta
\end{equation}
where the prime denotes differentiation. An alternative derivation of the above formula can be achieved by integration of the radial equation of motion \cite{raayai2019volume}.

Note that in contrast to the above approach, earlier theoretical models have been limited to determining the onset of fracture \cite{hutchens2016elastic,williams1965spherical,gent1991fracture,Williams1965}. Here we solve for the entire process by controlling the volume of the cavity.

To obtain a prediction of the expansion process, it remains to choose a constitutive relation. A natural choice is the neo-Hookean response, for which  $\bar W=(I_1-3)/2$ where for the present deformation, the first invariant of the left Cauchy-Green deformation tensor is $I_1=2\lambda_\theta^2+\lambda_\theta^{-4}$.  To account for strain stiffening that can appear due to  limiting chain extensibility, this relation can be extended using the Gent model \cite{gent1996new}
to write $\bar W=(J_m/2)\ln{(1-(I_1-3)/J_m)}$. Locking of the polymer network appears as a singularity when $I_1-3=J_m$. At the limit $J_m\to \infty$, the Gent model reduces to the neo-Hookean model. 

With the above formulation, for a given elastic response, prediction of the expansion process depends only on $\varphi$. As a first step we consider the neo-Hookean response, which is commonly adopted to model PDMS. For simplicity we choose $f(\lambda_a)=1$ in \eqref{UF}. In all figures (i.e Supplementary Information S4) the neo-Hookean prediction is represented by the black solid and dashed-dotted curves. Unless indicated otherwise, all of the results have been derived for samples with $B/A_0=15$, as in the experiments. 

It is already worth noting that all of the theoretical predictions have been made with a single value of the dimensionless measure of surface energy $\varphi=1$.

Comparing between the experimental curves and the neo-Hookean prediction for samples with $\mu\leq 48 \ \rm{[kPa]}$ reveals a notable agreement. Recall that no fitting parameters are used (other than setting $\varphi=1$). Focusing on the initial range of inflation (before the peak pressure) it is apparent that as the shear modulus increases, the experimental curves exhibit stiffening behaviour that is not captured by the neo-Hookean model, and that the increases in the peak pressures are followed by steeper gradients. The Gent model (as shown by the coloured curves) is able to capture the stiffening response within the elastic region and indicates that as the shear modulus increases (or equivalently the cross-linking increases) the elastic network can endure less stretching before locking (as indicated by the decreasing values of $J_m$). The good agreement between the experimental curves and the theoretical predictions in early stages of inflation before arriving at the peak pressure, confirms that within this range response is purely elastic. (Within this range the minimum energy solution is at the boundary $A = A_0$).  

To account for the fact that stiffening increases the energetic cost of fracturing as previously discussed \cite{gent1991fracture, lin2004cavity},  we have included the function  $f(\lambda_a)$ in \eqref{UF}. Based on the experimental results, at the neo-Hookean limit $J_m\to\infty$ this function should take the value $f(\lambda_a)=1$, while at the other limit, if the locking stretch is approached $(J_m\to I_1-3)$ the energy required to fracture becomes singular. Although different functional forms can be employed, in view of the Gent model, a natural choice is 
\begin{equation}
    \hat f(I_1)=\exp \left(\frac{J_m}{J_m-I_1+3}-1\right)
    \label{f1}
\end{equation}  
which, for generality, is written here in terms of the first invariant. For the present deformation pattern, \eqref{f1} can be specialized in terms of the circumferential stretch at the boundary, to write  $f(\lambda_a)=\hat f(2\lambda_a^2+\lambda_a^{-4})$. 

Now, without any additional fitting parameters, we apply the above formulation along with \eqref{f1} to the entire range of expansion. As observed, this simple model is able to capture the response over the entire range of shear moduli (Supplementary Information S4). Perhaps the most remarkable result is that it brings further insight into the physics of the problem by giving rise to first order transitions between metastable states (shown by the dashed-dotted coloured lines). Note that the metastable path for fracture corresponds to a local minimum in the energy $U$, that appears while the global minimum corresponds to the purely elastic path with $A = A_0$. At a critical expansion, the purely elastic solution is no longer a global minimum and fracture initiates $(A>A_0)$ (representative curves for the energy landscape are shown in the Supplementary Information S5).  As observed, these transitions become more pronounced for stiffer materials and explain the sudden drops in pressure. Since a first order transition requires excess energy, the onset of fracture in stiffer materials varies significantly among the different tests. Nonetheless, the curves eventually unite and resume similar trends.  

With the new understanding of this phenomenon, we can now return to answer the original question:  \textit{Is it cavitation or fracture?} To do so, we first examine the propagation of the fracture front $(A)$ throughout the expansion process on a phase diagram, as shown in Fig. \ref{UF}b. In this plane, travelling in the horizontal direction corresponds to purely elastic expansion, while processes that involve only fracture without cavity expansion travel in the vertical direction. As observed, initially all curves travel along the horizontal line in a purely elastic process with $A=A_0$. At a critical expansion $(\lambda_c)$, departure from the horizontal path is observed and becomes more abrupt as stiffening increases. Eventually all curves follow a path that is driven by the simultaneous action of both mechanisms. For materials with $J_m=3.8$ the metastable path departs earlier (i.e. $\lambda<\lambda_c$), a first order transition is then observed between the purely elastic expansion and a finitely fractured state. The transition corresponds to fracturing without any elastic effects, but then cavity expansion is resumed. Overall, throughout the process, expansion can abruptly switch between different mechanisms. 

Now that it is clear that both mechanisms can be simultaneously activated, another means for determining which mechanism is more dominant, is by considering the fraction of energy invested in each. 
In Fig. \ref{UF}c we plot the fraction of energy that has been invested in fracturing (for $J_m=3.8$ and $J_m\to \infty$). We find that within a large range of expansions $\lambda<4$ the fraction of energy invested in fracture does not exceed $60\%$. It is only when the effective fracture radius approaches the external boundary of the body that fracture engulfs, as observed for the curves with $B/A_0=15$.

Often, the debate between cavitation and fracture  revolves around the fact that fracture is associated with an initial length scale \cite{Williams1965,gent1991fracture} (i.e. the size of an initial defect or a notch), while cavitation theories exclude that information by considering the limit in which the initial defect is sufficiently small compared to the size of the body. In this case $B/A_0\to\infty$, or equivalently $\lambda_b=1$ (as seen from the incompressibility constraint). 
As shown in Supplementary Information (S5), the present model is practically insensitive to the ratio $B/A_0$ within a large range of expansion ratios.  Hence, the  formulation in equations \eqref{UE}-\eqref{pE} can be recast in an even  simpler form by substituting $\lambda_b=1$.

Finally, we recall that all of the theoretical predictions have been calculated using $\varphi=1$. This is a surprising result given the observed sensitivity to the shear modulus $(\mu)$ and that the value of $\varphi$ directly determines both the point of departure from the purely elastic response and the trend of the curve at large strains. Nonetheless,  $\varphi=1$ is shown to provide good agreement with experiments in all cases. Hence, for the PDMS samples used in this work $G_c=\mu A_0$ is proportional to the shear modulus. With $A_0\approx 1 \ \rm{[mm]}$ we find that $G_c$ is in the range of $25-250~ \rm{[J/m]}$. Since in the present formulation $G_c$ is calculated per effective unit area $S=2\pi(A^2-A_0^2)$, which can be thought of as the minimal surface area needed to advance the fracture front from $A_0$ to $A$, a geometric correction should be included in comparing the present value of $G_c$ with values measured via alternative methods.  

The energy release rate (per effective unit area) is obtained by directly differentiating \eqref{UE} with $\lambda_b=1$, which takes the simple form $\dot U_E=-{\partial U_E}/{\partial S}=\mu A \bar W(\lambda_a)$. To obtain the critical value at onset of fracture (when $A=A_0$), we substitute in this relation the critical stretch  $\lambda_c$, as determined from the energy minimization. It is observed from our experiments (Figs. \ref{curves} and more in Supplementary Information S4), that at the neo-Hookean limit $\lambda_c\approx1.55$ which leads to $\dot U_E=\mu A_0(=G_c)$. For increasing material stiffening, fracturing requires significantly higher levels of $\dot U_E$, as shown in Supplementary Information (S5).

Several studies have considered the relationship between the critical fracture energy release rate $(G_c)$ and the shear modulus $(\mu)$. The  classical model of Lake and Thomas \cite{lake1967strength} considers chain scission as the dominant mechanism of fracture and predicts that critical energy release rate follows a scaling of the form $G_c \propto \mu^{-1/2}$. Recently, Mao and Anand \cite{mao_fracture_2018} suggested that if fracture occurs by both cross-link stretching and scission, the scaling is $G_c \propto \mu^{1/2}$. Based on recent experiments and in accord with the classical models, Hutchens et al. \cite{hutchens2016elastic}  proposed a phenomenological power law model in the form $G_c \propto \mu^{n}$.   They report a range of the powers $n \approx 0.08-3$ for a few representative polymer systems. Noting the variations in the power $n$ in the previous literature, here, for PDMS samples we obtain  a linear dependence $(n=1)$.  Another key point to highlight is that our experiments clearly demonstrate a strong dependence of  the critical fracture energy release rate on the stiffening behaviour of the material ($J_m$), which has not been reported previously. 

To summarize, we have shown that in soft materials cavitation and fracture co-operate to drive failure. This observation is facilitated by a controlled experimental procedure of cavity expansion, that  allows us to both examine the cavity/crack patterns, and to measure the pressure-volume response with high precision. These tests are conducted on a family of PDMS samples with shear moduli in the range of $25-246$ [kPa]. A simple theory based on energetic arguments is shown to agree with all of the experiments while revealing that occurrence of abrupt pressure drops is a manifestation of a first order transition between metastable states, which may arise when material stiffening is significant. In exposing the existence of metastable states, this study explains the source of the significant dependence on initial imperfections, even in the absence of surface tension effects. Additionally, an effective measure of the critical fracture energy release rate is directly found from these experiments and is shown to be linearly proportional to the shear modulus, while exhibiting significant sensitivity to strain stiffening, which appears in these materials due to locking of the polymer chains at large stretches.  

Finally, while elucidating the basic mechanisms involved in formation of cavities/cracks in soft materials, the present work gives rise to a multitude of additional open questions. A natural next step is to extend this work to additional soft materials, to determine the microscopic mechanisms that lead to the dependence of fracture energy on stiffening of the polymer network, and to include surface tension effects that may become significant for smaller initial defects. It remains to be determined if similar interactions between cavitation and fracture appear in stiff materials, such as ductile metals.  Reminiscent of viscous fingering phenomena, our fracture patterns are characterized by various length scales and surface roughness. Additional work is needed to explain and perhaps, control, the formation of these elusive patterns. Finally, it is worth noting that the theoretical framework presented here, applies equivalently to situations in which tension is applied remotely. Although we have not been able to confirm experimentally that remote tension will result in similar expansion processes, nature has provided us with some examples (Supplementary Information S1).

\bibliography{apssamp}

\newpage
\onecolumngrid

{
\baselineskip = 18pt

\begin{center}
\textbf{\large Supplementary Information for the Manuscript}

\textbf{\large ``The Intimate Relationship between Cavitation and Fracture"}

Shabnam Raayai-Ardakani, Darla Rachelle Earl and Tal Cohen
\end{center}
}

\maketitle

\setcounter{equation}{0}
\setcounter{figure}{0}
\setcounter{table}{0}
\setcounter{section}{0}
\setcounter{subsection}{0}
\setcounter{page}{1}
\makeatletter
\renewcommand{\theequation}{S\arabic{equation}}
\renewcommand{\thefigure}{S\arabic{section}.\arabic{figure}}
\renewcommand{\thesection}{S\arabic{section}}
\renewcommand{\thesubsection}{S\arabic{subsection}}

\setcounter{section}{0}
\makeatletter
\renewcommand{\thesection}{S\arabic{section}}
\renewcommand{\thesubsection}{S\arabic{subsection}}

\label{S1}
\setcounter{figure}{0}

\section{Methods}
\label{mechod}

\textbf{Experimental Setup and Procedure} - Our custom designed Volume Controlled Cavity Expansion setup consists of a set of two attachments to an Instron universal testing machine as discussed by Raayai-Ardakani et al. \cite{raayai2019volume}. In summary, the stationary part is connected to the base of the Instron$^\circledR$ machine and holds a horizontal platform where the flange of the syringe is kept fixed. The moving attachment is connected to the load cell and moves with the Instron$^\circledR$ crosshead. The plunger of the syringe is fixed to this attachment and its movement is controlled and recorded with the movement of the crosshead.  In all the experiments 3mL BD syringes, and G19 and G21 needles were used. Any incompressible fluid (immiscible in the specimen) can be used in these volume-controlled experiments. The reported tests were conducted using Glycerol. A manual vertical translation stage (Optics Focus) is fixed to the base of the machine and is used to move the sample up and down to insert the fixed needle into the sample. Thus only the bottom of the sample is kept stationary, and the sample it is not confined on any of the other five faces. 

In the experiments, we impose the displacement of the crosshead (i.e. injected volume) and measure the force experienced by the syringe piston. The load cell of the Instron machine allows for accurate measurement of the force ($F$) applied to the plunger. The measured force is then used to find the pressure inside the cavity (calibration protocol is discussed by Raayai-Ardakani et al. \cite{raayai2019volume}). The injection rates did not exceed $5 \ \mu L/s$ to ensure quasi-static response.

\textbf{Sample Preparation} - We use soft PDMS (Polydimethylsiloxane) rubber samples (Sylgard 184, Dow Corning). To prepare the samples, the two parts were mixed at various base:cross-linking agent ratios and de-gassed using a vacuum pump. Afterwards, the samples were poured into molds of rectangular cuboid shape with a square cross sectional area of $30 \times 30 \ \rm mm^2$ and a height of $50 \ \rm mm$. Molds were filled with about $\sim 35 \ \rm mm$ of the mixture resulting in average heights of about $35 \ \rm mm$ after the end of de-gassing process. All the samples were cured in a $40^{\circ} \rm C$ oven for 72 hours. Afterwards samples were retrieved from the oven and left to cool to room temperature. All the experiments were performed at room temperature.

\textbf{Measurement of the Shear Modulus} - Shear modulus of the various PDMS samples of $\mu < 61 \ \rm [kPa]$ along with the initial effective radius of the cavities have been measured using the Volume Controlled Cavity Expansion technique as discussed by Raayai-Ardakani et al. \cite{raayai2019volume} and confirmed by tensile tests and previous measurements as presented earlier. The Shear modulus of the rest of the samples were measured using tensile tests and then used to evaluate the initial effective radius of the cavity based on the Gent constitutive model at $\lambda<1.35$. The value of $J_m$ for samples of $\mu > 50 \ \rm [kPa]$ have been found through trial and error.

\textbf{Casting and Visualization of Cavities} - After the completion of each experiment, the working fluid inside the cavities were completely drained. Then following a casting technique first proposed by Williams and Lofgren \cite{williams1988nest, tschinkel2004nest} for nest casting of ant species in Florida, we re-fill the emptied cavities with liquid polyvinyl siloxane (PVS - Zhermack Thechnical, Elite $^\circledR$ double 22). PVS comes in two liquid parts that are mixed in a 1:1 ratio and has a 10 minute working time. The liquid mixture is then drawn into a syringe and is inserted into the cavity with a needle and left to cure for 20 minutes at room temperature. The cured polymer inside the transparent PDMS samples freezes the shapes of the cavities and allows for easier visualization of the fracture patterns on the cavity walls. Ultimately, filled cavities of samples were photographed using a Nikon Micro-Nikkor lens (105 mm, f/2.8). 

\newpage 

\section{Natural Example of Crack Propagation due to Remote Tension}

\begin{figure}[!ht]
\centering
\includegraphics[width=0.45\textwidth]{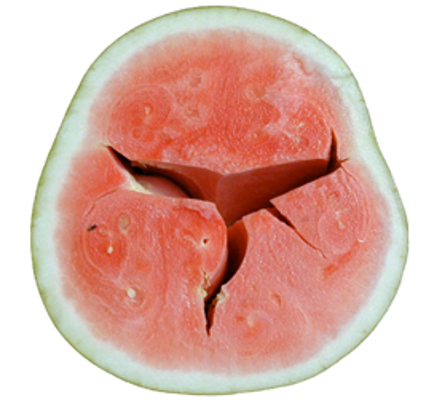}
\caption{Internal stresses generated by the confinement of the stiff crust during disruptive weather patterns, can lead to formation of cavity/crack in watermelon and other crop. This phenomenon is also referred to as hollow heart. This image is taken from: https://www.watermelon.org/Retailers/Hollow-Heart, and is not subject to copyright.}
\label{hollow}
\end{figure}

\newpage
\section{Additional Cavity Images for PDMS samples with Various Shear Moduli}
\label{AddCav}

\setcounter{figure}{0}
\begin{figure}[!ht]
\centering
\includegraphics[width=0.9\textwidth]{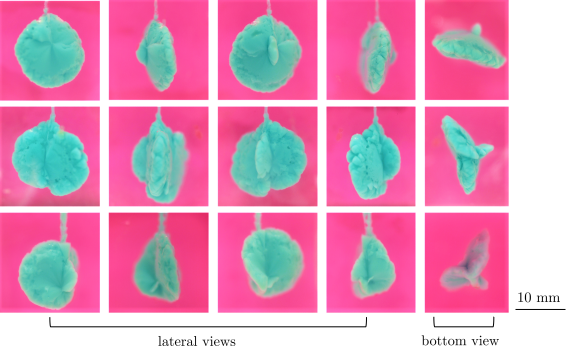}
\caption{Lateral and bottom views of the expanded cavities/cracks in transparent PDMS samples with $\mu=25$ [kPa]. Each row corresponds to a different test.}
\label{mu25}
\end{figure}

\begin{figure}[!ht]
\centering
\includegraphics[width=0.9\textwidth]{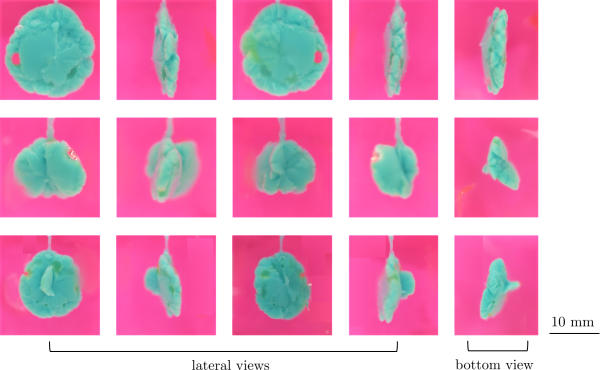}
\caption{Lateral and bottom views of the expanded cavities/cracks in transparent PDMS samples with $\mu=31$ [kPa]. Each row corresponds to a different test.}
\label{mu31}
\end{figure}

\begin{figure}[!ht]
\centering
\includegraphics[width=0.9\textwidth]{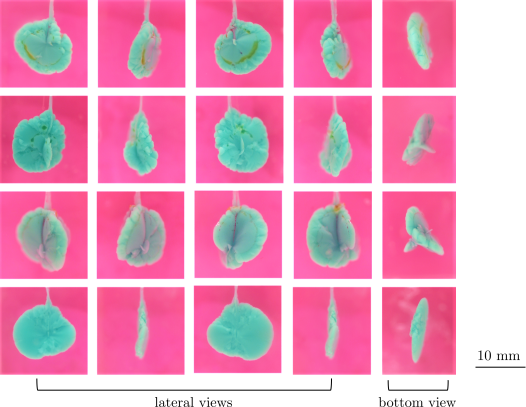}
\caption{Lateral and bottom views of the expanded cavities/cracks in transparent PDMS samples with $\mu=48$ [kPa]. Each row corresponds to a different test.}
\label{mu48}
\end{figure}

\begin{figure}[!ht]
\centering
\includegraphics[width=0.9\textwidth]{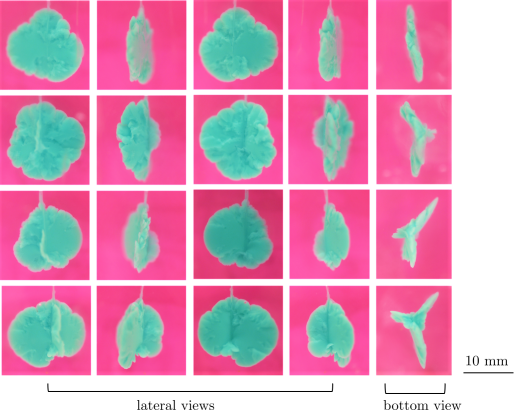}
\caption{Lateral and bottom views of the expanded cavities/cracks in transparent PDMS samples with $\mu=61$ [kPa]. Each row corresponds to a different test.}
\label{mu61}
\end{figure}

\begin{figure}[!ht]
\centering
\includegraphics[width=0.9\textwidth]{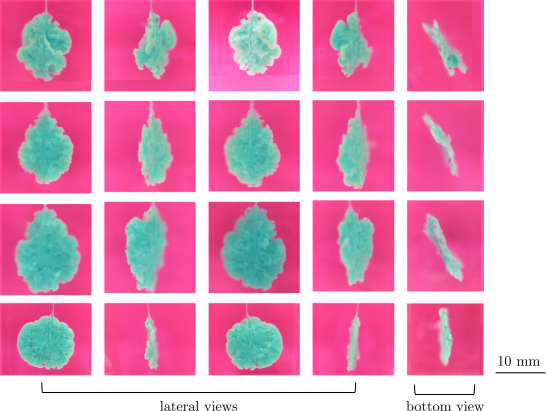}
\caption{Lateral and bottom views of the expanded cavities/cracks in transparent PDMS samples with $\mu=117$ [kPa]. Each row corresponds to a different test.}
\label{mu117}
\end{figure}

\newpage 
\ \ \ 
\newpage
\ \ \ 
\newpage
\ \ \ 
\newpage
\ \ \ 
\newpage

\section{Additional Curves of Normalized Pressure as a Function of the Stretch for PDMS Samples of Various Shear Moduli}
\label{AddCurves}
\setcounter{figure}{0}

\begin{figure}[!ht]
\includegraphics[width=0.95\textwidth]{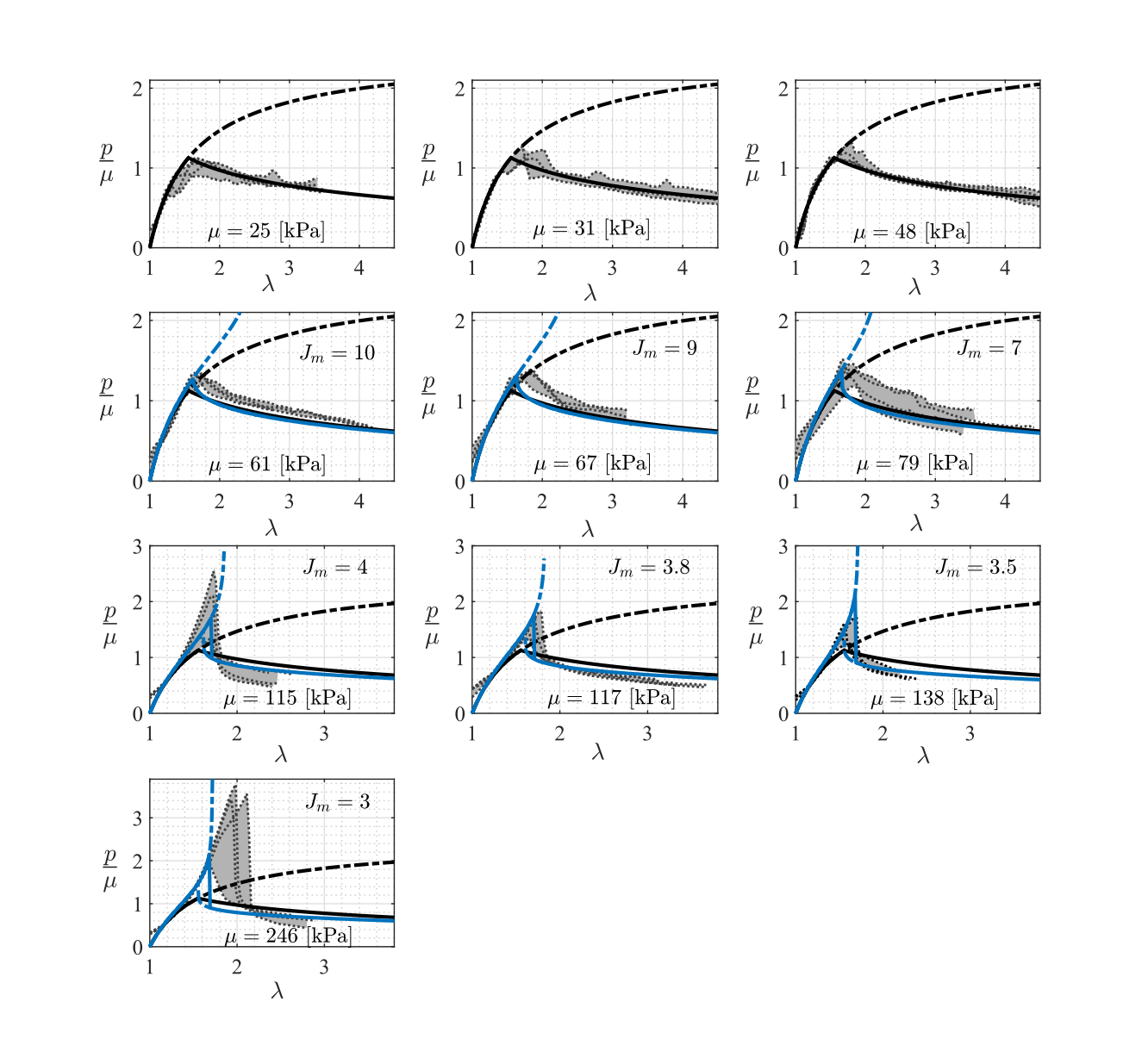}
\caption{Normalized Pressure, measured inside cavities of PDMS samples with different elastic moduli shown as a function of the effective stretch. Experimental curves are shown by the thin dotted lines. At least three tests were conducted for each material. The solid and dashed-dotted lines represent the predicted response using the present model, with the dashed-dotted lines representing metastable paths. All theoretical curves are obtained with $B/A_0=15$.}
\label{all_array}
\end{figure}

\ \ \ 
\newpage

\section{Sensitivity Analysis}
\label{BA}
\setcounter{figure}{0}

\begin{figure*}[!ht]
\centering
\includegraphics[width=0.8\textwidth]{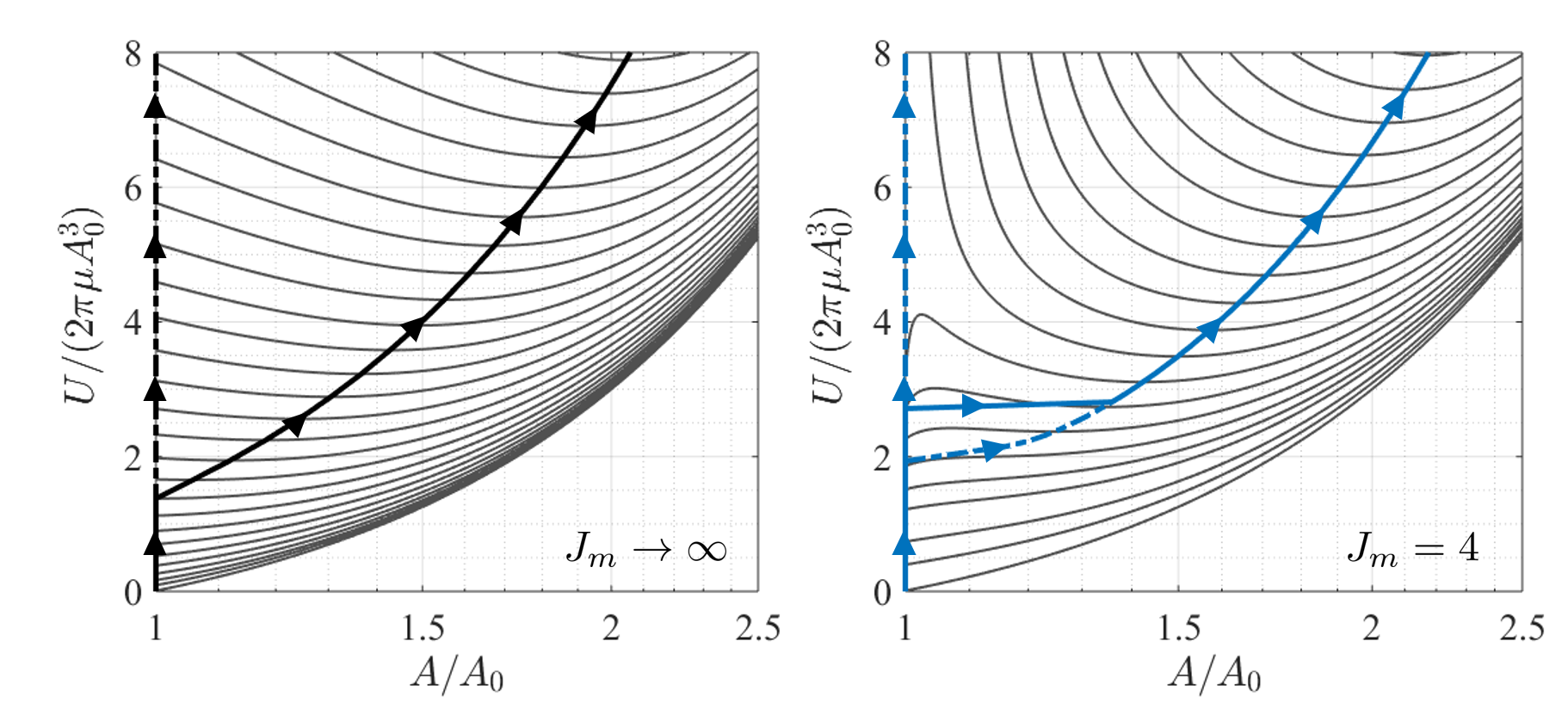}
\caption{Normalized energy landscape for varying locations of the fracture boundary $(A/A_0)$ and for different expansions shown by the full grey lines for different vales of $J_m$. The thick solid and dashed-dotted lines represent the predicted response using the present model, with the dashed-dotted lines representing metastable paths. All theoretical curves are obtained with $B/A_0=15$. }
\label{meta}
\end{figure*}

\begin{figure*}[!ht]
\centering
\includegraphics[width=0.5\textwidth]{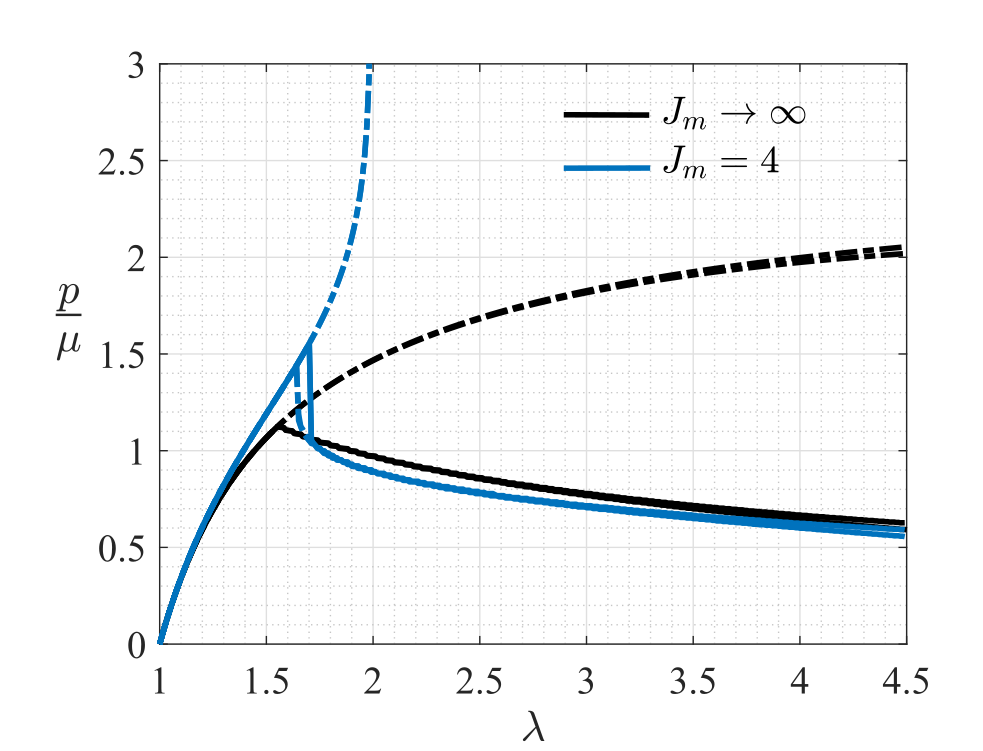}
\caption{Normalized cavity pressure shown as a function of the effective stress. The solid and dashed-dotted lines represent the predicted response using the present model, with the dashed-dotted lines representing metastable paths. Curves for $B/A_0=15$ and $B/A_0\to\infty$ are indistinguishable within a relevant range of expansion $\lambda=a_0/A$.}
\label{BA_sens}
\end{figure*}

\begin{figure*}[!ht]
\centering
\includegraphics[width=0.5\textwidth]{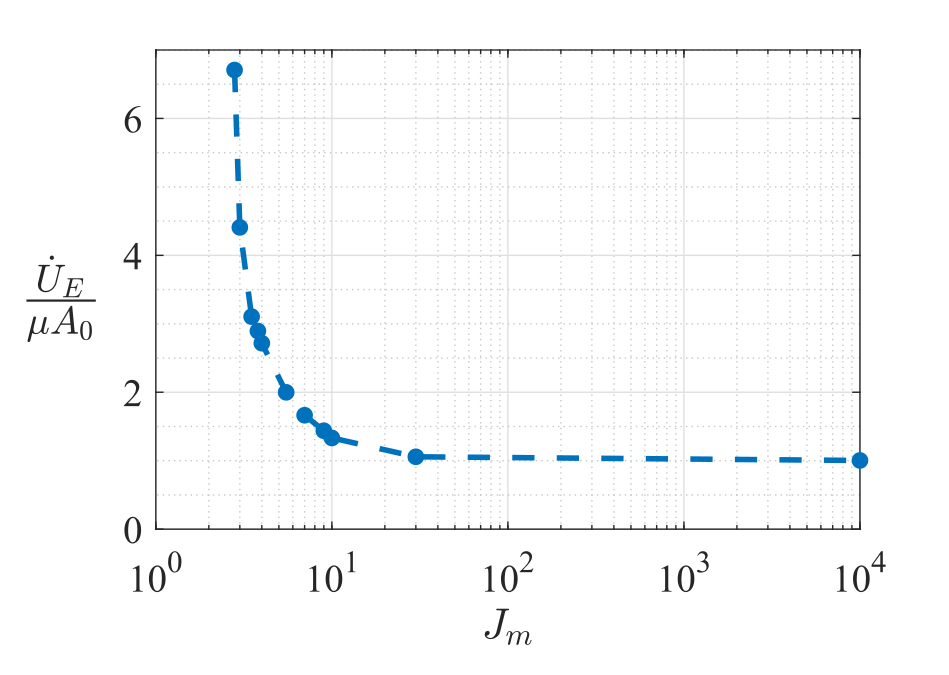}
\caption{Variations in the rate of energy release per unit area (normalized by shear modulus and initial cavity radius) as a function of the stiffening of the material.}
\label{dUE}
\end{figure*}

\ \ \ 
\newpage

\end{document}